\documentclass[pra,letterpaper,onecolumn]{revtex4} 
\usepackage{amsbsy,amssymb,amsmath,bm,bbold} 
\usepackage{graphicx,color,epsfig,rotate} 
\usepackage{fancyhdr} 
\usepackage{epstopdf}
\usepackage{graphicx}

\usepackage{float}
\usepackage[colorlinks=true,linkcolor=blue,citecolor=blue,urlcolor=blue]{hyperref}

\begin{document} 
	
	\title{Exploring Unconventional Features Of Light Dynamics
In Aubrey-Andr\'e-Harper Model Based Quasi-periodic Optical Lattices}
	
	\author{Suman Dey$^1$}
	\author{Nikhil Ranjan Das$^1$}
	
	\author{Somnath Ghosh$^2$$^*$}
	
	\affiliation{$^1$ Institute of Radio Physics and Electronics,University of Calcutta, West Bengal, India}
	\affiliation{$^2$ Unconventional Photonics Laboratory, Department of Physics, Indian Institute of Technology Jodhpur, Rajasthan 342037, India}

	\email{somiit@rediffmail.com}
	
		\maketitle %
	
\section{Abstract}
We report an Aubrey-Andr\'e-Harper (AAH) model based quasi-periodic lossless evanescently coupled waveguide lattice to study the unconventional physics of light localization. We present an exclusive methodical analysis of the band-topology of a tight-binding discrete lattice and accordingly study the modal characteristics to reveal the fact that a higher value of quasi-periodic modulation strength is imperative for observing a signature of fully localized light states having higher eigenenergy. This analytical concept has numerically been implemented in the proposed topological lattice to achieve light localization, where we have shown that the supported states not only depend on topological parameters, but also on the specific location of excitation which is supported by the violation of bulk-edge correspondence due to quasi-periodicity. Furthermore, we have investigated a unique effect of the presence of disorder on light localization phenomenon, where it has been reported that the presence of off-diagonal disorder, which is otherwise detrimental, favours light localization in the proposed structure due to topological protection. The findings indeed have the potential to open up a fertile platform to manipulate light in topologically aided passive photonic devices.

	\section{Introduction}
	
		Recent expeditious advancements in photonics in the form of fundamental understandings, translational research and innovations clearly establish that topological photonics \cite{ozawa19} is likely to be one of the emerging domain for next-generation photonics technology. Even though topology is a branch of mathematics, it has been rigorously and efficiently exploited in condensed matter domain of physics during the past two decades. As a hallmark, topological insulator \cite{shankar18} has preserved bulk states whereas the counterpart edge states of it are perfectly conducting. In this context, on the basis of the idea of topological insulator, Halden and Raghu first proposed the photonic analog  of it \cite{haldane08}, which opened a huge scope for potential investigation of the unconventional states of light in topological photonic structures. A topological structure potentially supports its protected states which are robust against any small and unwanted external perturbations. Specifically, owing to the tuning of topological parameters, a transition from  trivial to non-trivial  topological phase \cite{lu14} is possible in the same photonic structure which essentially supports the photonic  edge states, analogous to quantum Hall  edge states in condensed matter \cite{raghu08} physics. Moreover, numerous topological structures have been explored to implement them as photonic passive/active devices \cite{rider19,wu17}. Among these structures photonic quasicrystals ($PhQC$) have been identified to have a direct correspondence with one of  the topology-driven models \cite{liu15}. The $PhQC$ itself is an aperiodic structure with long-range order \cite{vardeny13} and can be thought of as projection of a periodic structure of higher dimension. $1$D Fibonacci quasicrystal is the best example of  projection of a 2D lattice on a line \cite{janot92}. Even though $PhQC$ has no translational symmetry, it shows unconventional Bragg diffraction \cite{vardeny13}. $AAH$ model based $PhQC$ shows some interesting topological properties which can be related to condensed matter physics \cite{kraus12,guo_18}. Incommensurate $AAH$ lattice without having adiabatic phase term also has been  reported \cite{lahini09} . 
		
However, except some aforementioned  basic topological features of quasi-periodic lattices, a detailed physical investigation in this context is lacking. For example, topological phase transition feature of $AAH$ lattice without having adiabatic phase term has never been explored in terms of band-topology analysis. Moreover, modal characteristics, light localization behaviour and its simultaneous dependency on topological parameters with other system parameters are yet to be reveal. Hence, in this work, we present a detail investigation in the context of judicious manipulation of light-states in an $AAH$ model based physical quasi-photonic structure. Such investigations may convey a new physical insight in the context of light manipulation in topologically controlled quasi-periodic lattice structures. 
		
In this paper, the topological behavior of 1D-$AAH$ lattice is derived through the systematic analysis of the band topology. We show that the signature of  strong localization of higher energy eigenstates depends on topological modulation strength ($\lambda$) of quasi-potential due to their inherent modal characteristics in $AAH$ lattice. Also, we implement the $AAH$ discrete lattice model in a lossless evanescently coupled dielectric waveguide lattice and report that the light localization behavior for left edge, center and right edge  excitations in the structure, are distinctly different and governed by a fixed topological parameter.  Moreover, the effect of topology on the localized light state in presence of deliberate off-diagonal disorder in the structure is also investigated. 
	\section{Analytical implementation using discrete tight binding lattice approximation}
	 We consider a 1D tight binding discrete lattice with next nearest neighbor hopping and its supporting eigenvalue equation in which on-site potential is varied in space can be described as \cite{aubry80} 
	\begin{equation}
	C(\Psi_{n+1}+\Psi_{n-1})+\lambda \cos(Q n)\Psi_{n}=E_{n}\Psi_{n}
	\label{tightbinding}
	\end{equation}
	Here, $C$ is the tunneling rate or hopping amplitude, $\Psi_{n}$ is the wavefunction of site $n$, $\lambda$ is the potential modulation strength, and $Q$ controls the periodicity of the modulation. For  onsite commensurate  and incommensurate modulation within lattice the $Q$ values follow the relations  $Q/2\pi=5/3$ (3-periodic) and $Q/2\pi=(\sqrt{5}+1)/2$ (quasi-periodic) respectively, where $(\sqrt{5}+1)/2$ is the golden ratio which is also known as irrational Diophantine number. In Eq. \eqref{tightbinding}, the terms associated with $\lambda$ are diagonal elements, which consist the quasi-periodic onsite potential. Whereas the terms associated with the coefficient $C$ are off-diagonal elements, which represents the coupling term ($C$) which is called tunneling coefficient. Here, we consider $C=1$ which defines homogeneous tunneling between next nearest neighbor lattice sites ($n-1$ and $n+1$ sites with $n$ sites).
	\begin{figure}[t]
		\centering
		\includegraphics[width=8.5cm]{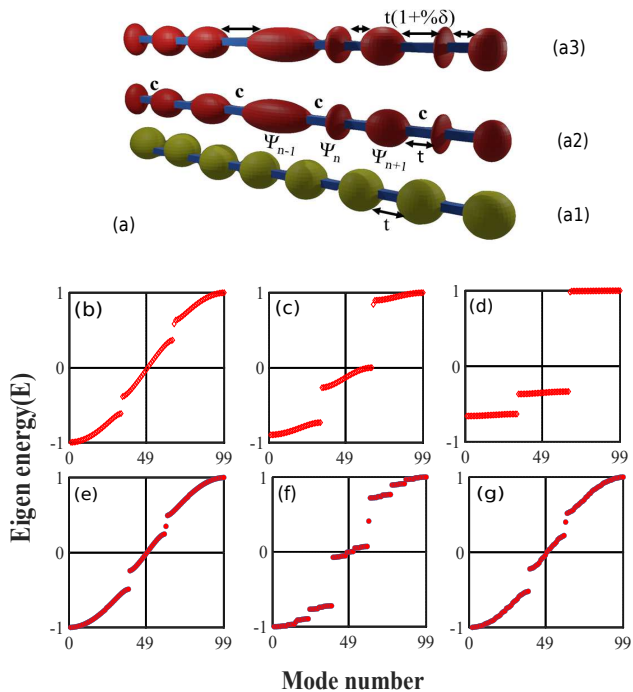}
	\caption{(Colour online) (a) The schematic representation of tight binding discrete lattices: (a1) The periodic commensurate tight binding lattice; (a2) Incommensurate quasi-periodic lattice without disorder; (a3) Incommensurate quasi-periodic lattice in presence of off-diagonal disorder where $t$ is the separation between two consecutive sites in the corresponding lattice which can vary randomly. (b)-(g)  Numerically calculated normalized spectrum from Eq. \eqref{tightbinding} with consideration $C$ as 1 and $n$ as 99. (b)-(d) Spectrum for commensurate modulated lattice. (e)-(g) Spectrum for incommensurate modulated lattice. The band plots are for different  potential modulation strengths, for example (b) and (e) $\lambda=0.5$, (c) and (f) $\lambda=2$, (d) and (g) $\lambda=6.5$. The incommensurate (quasi-periodic) lattice shows the trivial to non-trivial phase transition after $\lambda=2$. }
	\label{fig:bandplot}
  \end{figure}
  \subsection{Band-topology analysis}
	In the proposed schematic shown in Fig. \ref{fig:bandplot}(a), the shape of the spheres represent onsite potential variation. The periodic commensurate tight binding lattice is shown in Fig. \ref{fig:bandplot}(a1) where shapes of all spheres are identical. Fig. \ref{fig:bandplot}(a2) represents incommensurate lattice without disorder  where shapes of spheres are chosen to vary as quasi-periodically and Fig. \ref{fig:bandplot}(a3) represents  incommensurate lattice in presence of off-diagonal disorder where $t$ is the separation between two consecutive sites in the corresponding lattice which can vary randomly. As we know that for 1D tight binding lattice in absence of any adiabatic phase term \cite{liu15}, eigenenergy vs mode number plot is defined as band plot \cite{2008}. Besides, it is known that slope of band is minimum at topological phase transition point \cite{liu15} for any topological model, while before and after topological transition, slope gradually increases. So, on the basis of that a direct comparison between spectrum of commensurate and incommensurate lattices has been shown in Figs. 1(b)-(g) for the model as shown in Fig. \ref{fig:bandplot}(a2). In Figs. \ref{fig:bandplot}(b)-\ref{fig:bandplot}(d), as modulation strength ($\lambda$) increases from 0.5 to 6.5, the slope of the discrete band becomes flat  after $\lambda=2$. However, in Figs. \ref{fig:bandplot}(e)-\ref{fig:bandplot}(g) not only the number of quasi band-gaps increases with increasing $\lambda$, slope of the discrete bands also becomes flat at $\lambda=2$. Hereafter the band slope becomes steeper and at $\lambda=6.5$ the band slope again looks exactly similar to that of $\lambda=0.5$. Thus, by  direct comparison of spectra between periodic (commensurate) and quasi-periodic (incommensurate) modulated tight binding lattices, trivial to non-trivial phase transition of the band can be witnessed  only  for incommensurate  modulation after $\lambda=2$. Hence it clearly reveals that potential modulation strength ($\lambda$) is the topological parameter for $AAH$ model with respect to the band-topology. This methodical analysis of band-topology is very helpful to determine the topological parameter of any 1D aperiodic structure which inherits topological property.
	\begin{figure*}[t]
		\centering
		\includegraphics[width=16cm]{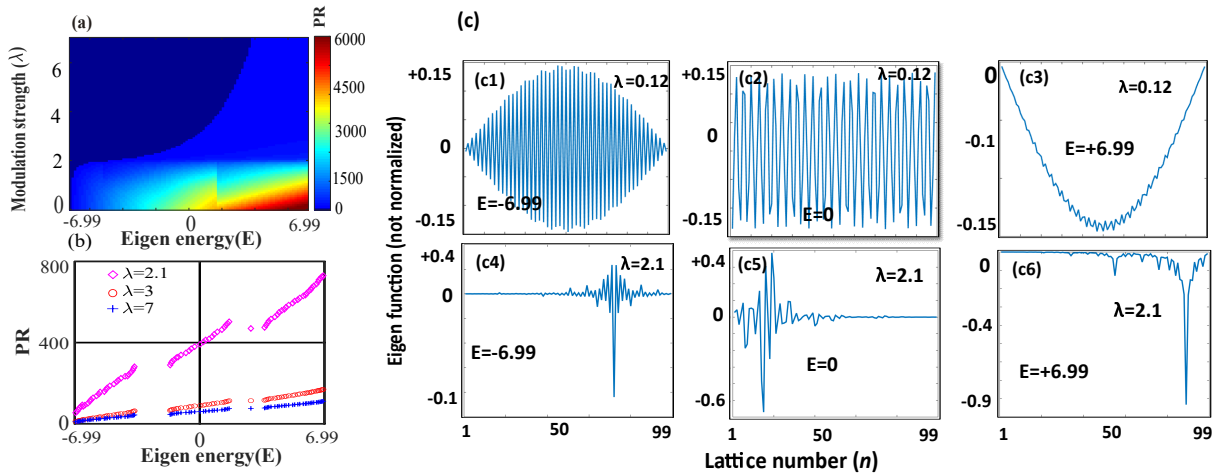}
		\caption{(Colour online) (a),(b) Variation of PR with eigenenergy values of different modulation strengths ($\lambda$) for a direct comparative study of localization behaviour with consideration 99 lattices ($n$). (c) The distribution of three eigenfunctions corresponding to lower most ($E=-6.99$), band center ($E=0$) and higher most ($E=+6.99$) eigenenergies  for before and after topological phase transition: (c1),(c4) Eigenfunctions ($\Psi$) corresponding to lowest eigenenergy ($E=-6.99$) for before and after topological phase transition respectively. (c2),(c5) Eigenfunctions ($\Psi$) corresponding to band-center eigenenergy ($E=0$) for before and after topological phase transition respectively. (c3),(c6) Eigenfunctions ($\Psi$) corresponding to highest eigenenergy ($E=+6.99$) for before and after topological phase transition respectively. }
		\label{fig:PR}
	\end{figure*}
	
\subsection{Exploring modal characteristics and their behaviour towards localization}

After analyzing the exclusive band-topology, we explore the topological dependency of the modal characteristics and their localization behaviour in the proposed AAH lattice [as shown in Fig. \ref{fig:bandplot}(a2)] for the span of total eigenenergy $-6.99\le E\le6.99$. For the considered parameters of the AAH lattice, this maximum eigenenergy span has been predicted from Eq. \eqref{tightbinding}. Now to investigate the localization phenomena in the modal behaviors, we calculate the eigenfunctions from Eq. \eqref{tightbinding}, where, the corresponding participation ratio (PR) of each of the modes have been calculated as $\text{PR}=\textstyle (\sum_n|\Psi_{n}|^{2})^2/\sum_n|\Psi_{n}|^{4}$ ($n=99$) \cite{lahini09}. 

Figure \ref{fig:PR}(a) shows PR (color-axis) of the modes corresponding to the entire eigenenergy span of the proposed AAH lattice for $0\le\lambda\le7$. Here the span of lambda has been chosen to show the distribution of PR at both before ($\lambda<2$) and after ($\lambda>2$) the topological phase transition. Here, the PR of a particular mode describes its localization behavior, where a eigenstate having smaller PR  is more localized than a eigenstate having  higher PR. From the distribution of PR in Fig. \ref{fig:PR}(a), it is evident that the eigenstate corresponding to lower most eigenenergy ($E=-6.99$) becomes strongly localized with small PR when the value of $\lambda$ just exceeds the value 2. However, other states corresponding to comparably higher eigenenergies are not strongly localized, as they have very high PR-values.

To visualize this explicitly, the localization behaviour has been shown in Fig. \ref{fig:PR}(b) for three specifically chosen modulation strengths ($\lambda$) after the topological phase transition ($\lambda>2$). Here, it is evident that for a fixed $\lambda= 2.1$, PR increases with respect to the increasing eigenenergy. Here, $\lambda= 2.1$ has been chosen to consider the situation immediately after the topological phase transition. Now, it has also been observed in Fig. \ref{fig:PR}(b) that if the modulation strength is further increased (e.g. $\lambda=3,7$), PR reduces significantly for each of the eigenstates. Hence, strong localization of higher energy states can be realized if the modulation strength is further increased ($\lambda>>2$). This special localization behaviour is achieved due to different characteristics of modes corresponding to different eigenenergies.

Now, to visualize the modal characteristics explicitly, we plot the distribution of three eigenfunctions (over the entire lattice) corresponding to lower most ($E=-6.99$), band center ($E=0$) and higher most ($E=+6.99$) eigenenergies in Fig. \ref{fig:PR}(c). Here, Fig. \ref{fig:PR}(c1)--(c3) represent the distribution of three extended modes before the topological phase transition, while considering $\lambda=0.12$. On the other hand, as can be seen in Fig. \ref{fig:PR}(c4)--(c6), localization has been observed in the distribution of three same modes for $\lambda=2.1$, i.e., immediately after the topological phase transition. Here, both the extended [Fig. \ref{fig:PR}(c1)] and localized [Fig. \ref{fig:PR}(c4)] modes associated to $E=-6.99$ look like a staggered mode \cite{2008} because its amplitude between two lattice points is changing from positive to negative and negative to positive simultaneously. However, for $E=+6.99$, both the extended [Fig. \ref{fig:PR}(c3)] and localized [Fig. \ref{fig:PR}(c6)] modes look like a flat phase mode \cite{2008} having totally negative phase. In between these two extreme cases corresponding to $E=-6.99$ and $E=+6.99$, both the extended [Fig. \ref{fig:PR}(c2)] and localized [Fig. \ref{fig:PR}(c5)] modes behave as a mixed phase mode for $E=0$. Hence, the localization behavior of different eigenstates after the topological transition is different. Such localization behavior of modes [as can be seen in Figs. \ref{fig:PR}(c4)--(c6)] after the topological phase transition can also be established form the PR distribution for the specific $\lambda=2.1$, as already shown in Fig. \ref{fig:PR}(b). Such topology based modal characteristics have also been investigated for different choices of $n$, where we can confirm that the localization phenomenon of a specific mode is unaltered with the increasing number of lattice sites ($n$).        

	\begin{figure*}[t]
		\centering
		\includegraphics[width=17cm]{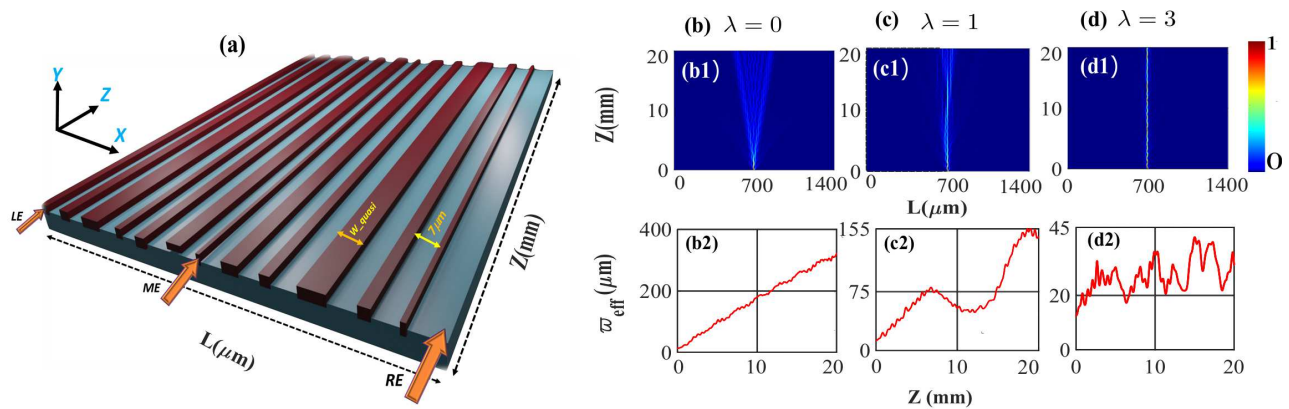}
		\caption{(Colour online) (a) A schematic of a lossless coupled waveguide lattice designed as $AAH$ model where regions in reddish colour  indicate the high refractive index($\xi$) and regions in light blue colour depict the low refractive index($\xi_{0}$) respectively and has a longitudinal length (Z) of 20 mm. (b)-(d) The beam propagation results of $ME$ for three different modulation strengths for e.g. $\lambda=0,1,3$ respectively where contour plots show the beam intensity variation during propagation (colour bar represents intensity,$\big|\psi\big|^{2}$) and other plots in the lower panel show variation of effective width ($\varpi_{eff}$) with $Z$. (b1),(b2) For $\lambda=0$, discrete diffraction behaviour is evident from contour plot and linearly increasing $\varpi_{eff}$ with $Z$ illustrates that the beam propagates with ballistic dynamics. (c1),(c2) For $\lambda=1$, similar type of plots show nearly discrete diffraction behaviour and a kind of increasing $\varpi_{eff}$ variation which  illustrate that the beam propagates with a transient ballistic state. (d1),(d2) For $\lambda=3$, plots illustrate that the beam propagates in the localized regime. Hence, there is a topological phase transition between $\lambda=1$ and $\lambda=3$ where the beam gets directly localized from ballistic state.
		}
		\label{fig:waveplot}
	\end{figure*}

\section{Design of an AAH-model based quasi-periodic waveguide lattice}

Now, to implement the analytical treatment associated with the tight-binding discrete AAH lattice [as shown in Fig. \ref{fig:bandplot}(a2)], we have designed an evanescently coupled and quasi-periodic waveguide lattice, as shown in Fig. \ref{fig:waveplot}(a). To construct the waveguide array, we have considered a base layer with the background refractive index $\xi_{0}$, where to consider the individual guides with high refractive index, a constant index difference $\Delta\xi$ has been maintained. Here, we have chosen $\xi_{0}=1.46$ and $\Delta\xi=0.001$. To establish the nature of incommensurate quasi-periodic onsite modulated potential of AAH model in the coupled waveguide structure, we have modulated the width of high-indexed layers (sites) quasi-periodically as $W_{quasi}=W_{high}+\lambda\cos(Qn)$. Here, the unmodulated width ($W_{high}$) of the high index sites have been chosen as $7\mu$m and $Q/2\pi=(\sqrt{5}+1)/2$. However, we have maintained the separation between two waveguides ($t$) as $7\mu$m.

Now, in the quasi-periodic refractive index profile of the proposed waveguide array, the high index sites ($n$ sites) are treated as onsite potential term [analogous to the diagonal terms associated with $\lambda$ in Eq. \eqref{tightbinding}]. Moreover, the unaltered separations between quasi-periodic high-indexed sites defines the homogeneous tunneling among the waveguides and the tunneled field from a high index site is coupled only with its two adjacent high index sites [analogous to the coupling between two next-nearest neighbors via the off-diagonal coefficient $C=1$ in Eq. \eqref{tightbinding}]. Now, the index contrast $\Delta\xi=0.001$ (i.e., the waveguides are initially very weakly guiding)  ensures that the overall system is a strongly coupled system and the tight-binding lattice approximation is valid as described by Eq. \eqref{tightbinding}. Thus, the proposed realistic waveguide array shown in Fig. \ref{fig:waveplot}(a) is analogically similar to the analytical model based on Fig. \ref{fig:bandplot}(a2) and the key features exhibited by tight binding coupled mode equation [Eq. \eqref{tightbinding}] can also be hosted by the proposed optical lattice. Using a state-of-the-art ultra-fast laser inscription technique \cite{psaila08,ghosh} and lithography, the proposed waveguide array can physically be realized.  
 
\subsection{Comparative analysis of localization behaviour of injected light beam in the optical lattice}
	 The light dynamics in the proposed optical lattice structure is governed by the Schrödinger like equation \cite{schwartz07}. In paraxial limit, it is described as
	\begin{equation}
	i\frac{\partial \psi(x,z)}{\partial z}=-\left[\frac{1}{2k}\frac{\partial^{2}}{\partial x^{2}}+\frac{k}{\xi_{0}}\Delta\xi(\lambda,x)\right]\psi(x,z)
	\label{wave_equation}	
	\end{equation} 
	where, $\psi(x,z)$ is the electric field amplitude of a monochromatic continuous wave (CW) optical beam. In general $E(x,z,t)=\text{Re}\left [\psi(x)\exp{(kz-\omega t)}\right]$, $kc_{0}=\xi_{0}\omega$ and $c_{0}$, $\omega$ are the velocity of light in vacuum and the angular frequency of the input beam respectively.
	We solve Eq. \eqref{wave_equation} by finite difference scalar beam propagation method (FD-BPM) \cite{pedrola15} for an input  CW Gaussian beam of wavelength 980 nm and FWHM 5 $\mu$m. The input beam, covering few lattice sites is assumed to be incident on the lattice unit located around the central region of the structure irrespective of the index of the local lattice unit. We call this as Middle-site Excitation ($ME$). Figures \ref{fig:waveplot}(b1)-\ref{fig:waveplot}(d1) show that the discrete diffraction leading to ballistic state occur if $\lambda<2$ and light localizes after $\lambda>2$ and detailed study has confirmed that the topological phase transition is occurred at $\lambda=2$ which is as predicted by band-topology in discrete $AAH$ lattice, shown in Figs. \ref{fig:bandplot}(e)-(g). To visualize these facts systematically and reveal the true nature of light dynamics, we calculate the effective width($\varpi_{eff}$) of the input Gaussian beam throughout the propagation. The effective width of a beam \cite{schwartz07} in 1D dielectric lattice geometry is defined as, $\varpi_{eff}(\mu m)=\textstyle (\int\big|\psi(x,z)\big|^{2} dx)^2/(\int\big|\psi(x,z)\big|^{4} dx)$. Figure \ref{fig:waveplot}(b2) illustrates the fact that for a perfectly ordered lattice ($\lambda=0$), $\varpi_{eff}$ increases linearly with propagation distance which confirms a proper ballistic dynamics of light. In Fig. \ref{fig:waveplot}(c2), $\varpi_{eff}$ increases with propagation distance but does not follow linear behavior because of asymmetric widths of high index sites as modulated. This may be referred as a transient ballistic mode. In Fig. \ref{fig:waveplot}(d2), the variation of effective width of the beam describes the localized state for $\lambda=3$ with some fluctuations which occur due to the quasi-periodic nature of the structure. Hence, we reveal the unconventional nature of light dynamics for $ME$ before and after topological phase transition in these special lattices.

\begin{figure}[t]
		\centering
		\includegraphics[width=8.5cm]{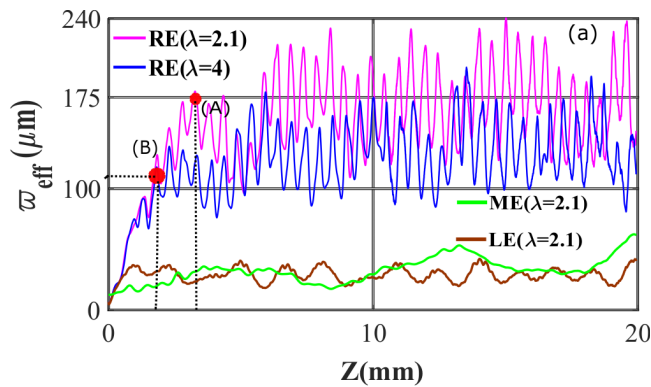}
		\caption{(Colour online) (a) Variation of the effective width of the propagating Gaussian beam for $LE$, $ME$ and $RE$ respectively, in the localized regime beyond $\lambda=2$ and chosen to be $ \lambda=2.1$ and $4$. Their localization characteristics are distinctly different compare to each other. The comparison of localization behaviour for $RE$ between $\lambda=2.1$ and $\lambda=4$ shows that as we increase $\lambda$, the beam becomes more localized  for $RE$.}
		\label{fig:edge_middle}
\end{figure}

	 To investigate the detailed light localization behavior, we inject the Gaussian beam near the left and the right edges of the structure which can be called as Left-edge Excitation ($LE$) and Right-edge Excitation ($RE$) respectively as shown in Fig \ref{fig:waveplot}(a). As a result, we show a direct comparison of light dynamics for a fixed topological parameter ($\lambda$) among $LE,ME$ and $RE$ in Fig. \ref{fig:edge_middle}(a). The beam covers  few lattice sites of the left edge for $LE$ and  the right edge for $RE$. In  Fig. \ref{fig:edge_middle}(a), we have shown that  for $LE$, strong localization is confirmed with a  small effective width  at $\lambda=2.1$ which just exceeds $\lambda=2$. Moreover, from Fig. \ref{fig:edge_middle}(a)  it is also evident that the beam for $ME$ has shown localized nature with higher values of effective width compare to $LE$ for the same value of modulation strength ($\lambda=2.1$). In contrast, the $RE$ depicts drastically different propagation characteristics for the same value of modulation strength ($\lambda=2.1$). In Fig. \ref{fig:edge_middle}(a)  for $RE$, the input beam at first follow  a transient ballistic behaviour up to the point marked by (A) and then it becomes localized with a relatively high $\varpi_{eff}$ compared to both $LE$ and $ME$. Hence, it does not exhibit strong localization. Therefore, it is established that all input beams for $LE$, $ME$ and $RE$ have been localized as soon as the modulation strength crosses the value 2 ($\lambda=2.1$). This behaviour reaffirms that the topological phase transition occur after $\lambda=2$ in quasi-periodic optical lattices where the localization behaviour of different excitations follow different trends. It solely depends on the location of the input excitation along with the choice of the $\lambda$. Besides, if we compare the effective width variation for $RE$ between two different cases for $\lambda=2.1$ and $\lambda=4$ in Fig. \ref{fig:edge_middle}(a) , it reveals that a transient ballistic state for $\lambda=4$ is followed up-to the point marked by (B) and it has the $\varpi_{eff}$ value smaller than that at the point (A). After reaching the point (B), the injected light  becomes localized with a relatively small value of effective width in comparison to $\lambda=2.1$ for $RE$. In this context, also from a direct comparison between Fig. \ref{fig:waveplot}(d2) and Fig. \ref{fig:edge_middle}(a), we confirm that the beam for $ME$ has been localized with a smaller effective width as we increase the modulation strength to $\lambda=3$. So, for the chosen values of $\lambda$, $\varpi_{eff}$ decreases  as $\lambda$ increases for both $ME$ and $RE$.  Moreover, from Fig. \ref{fig:edge_middle}(a) it is evident that for a chosen topological parameter, the light localization behaviour for the $LE, ME$ and $RE$ follows the prediction of the state dynamics corresponding to only specific lowest ($E=-6.99$), band-center ($E=0$) and highest ($E=+6.99$) eigenenergy  in  $AAH$ lattice, shown in Fig. \ref{fig:PR}(b). 
	 The physics behind these unconventional features is that the structure does not follow the ``bulk-edge correspondence" \cite{ozawa19}. We know that the bulk-edge correspondence is strictly valid for periodic topological photonic structure. However, in this context we may mention that the proposed structure has onsite quasi-periodic refractive index profile. So, due to lack of translational symmetry and long-range order, it has bulk
minibands in its band spectrum and there are mini gaps between successive bulk mini bands. Now, whenever the modulation strength ($\lambda$) of the onsite refractive index modulation has value less than 2 ($\lambda<<2$), the number of minigaps are small and there exist quasi-continuous bulk bands. So, the eigenstates are in extended in nature due to easy coupling between partially localized states of quasi-continuous bulk bands. As a result, we can state that the structure is in trivial phase and shows the transient ballistic nature. But when we increase the modulation strength, the number of mini gaps is increased (increasing the quasi-periodic nature of refractive index profile) and discontinuity between bulk bands is also increased. The probability of coupling of eigenstates of individual bulk bands has decreased and the span of the bulk bands has also become very narrow. So, they support only very few states in their continuous
bulk band which cannot spread over two or three lattice sites in the structure. This can be considered as a localized state and the structure has the trivial to non-trivial transition now. For each bulk band there is a localized eigenstate which is now topologically protected and so, there are many localized states which possess low to high eigenenergy. This feature can only be seen after $\lambda=2$ in $AAH$ model based quasi-periodic optical lattice. So, if the input excitation is placed anywhere in the structure in the non-trivial condition, some localized eigenstate will be excited corresponding to low to higher energy depending on position of excitation in the structure and topological parameter. In our paper, $\lambda$ was chosen to be $2.1$ for the comparison of topological localization among $LE$, $ME$ and $RE$. Besides, we can also emphasize that for the control of states of light  by topological parameter of this quasi-periodic lattice, the $ME$ has shown the topologically induced localized sate when the lattice is in non-trivial phase as shown in Fig. \ref{fig:waveplot}(d).
\subsection{The behaviour of light localization in presence of off-diagonal disorder in the optical lattice}
	  Till now, the unconventional behaviour of localization  of light in the proposed AAH optical lattice have been explored in absence of any disorder. Here, we have studied some unique features of light localization in the presence of deliberate disorder in the proposed optical lattice. It is well known that the off-diagonal disorder is detrimental for the localization of light. In this context, to explore the localization behaviour in presence of off-diagonal disorder in this special lattice for a fixed modulation strength ($\lambda=3$), we deliberately vary the separation between two consecutive waveguides randomly as $t=t(1+\% \delta)$  where, $\delta$ is the random number between $[-1,1]$ to introduce off-diagonal disorder in the proposed structure as shown in Fig. \ref{fig:waveplot}(a). 
	  \begin{figure}[t]
		\centering
		\includegraphics[width=8.5cm]{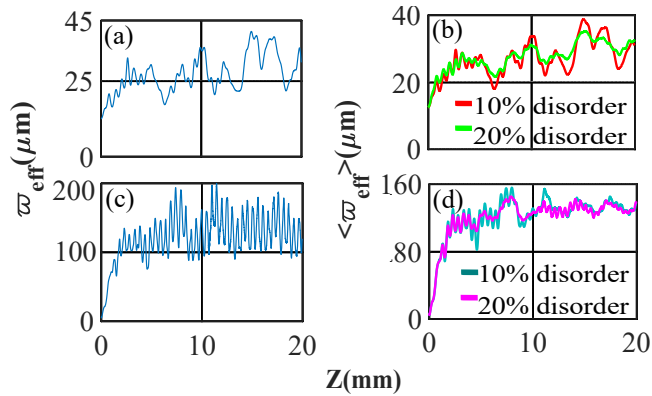}
		\caption{(Colour online) Effective width variation of a input beam  for $\lambda=3$. (a),(c) Without presence of any disorder for $ME$ and $RE$ respectively. (b),(d) In presence of off-diagonal disorder for $ME$ and $RE$ respectively.
		}
		\label{fig: disorder}
	\end{figure}
Figs. \ref{fig: disorder}(a), \ref{fig: disorder}(c) and Figs. \ref{fig: disorder}(b), \ref{fig: disorder}(d) show the effective width variation of the structures with and without off-diagonal disorder for the same input Gaussian beam respectively. If we compare Fig. \ref{fig: disorder}(a) with Fig. \ref{fig: disorder}(b) for $ME$ and Fig. \ref{fig: disorder}(c) with Fig. \ref{fig: disorder}(d) for $RE$, a closer look directly reveals that the fluctuations in $<\varpi_{eff}>$ variation in the localized regime has decreased in presence of off-diagonal disorder. Moreover, increasing the off-diagonal disorder strength from 10\% to 20\% does not have much effect on characteristics of localized  states of light.  Hence, the analysis of light dynamics in the designed optical lattice conclude that off-diagonal disorder favours the localization due to topology. The similar behaviour has also been observed for $LE$. This can be regarded as another unconventional feature of the light dynamics through these special lattices owing to the topological protection.
\section{Summery}

In summary, we have reported the topological dependence of unconventional light localization by using the framework of an AAH model based quasi-periodic waveguide lattice. We have analytically described the topological phase transition features of $AAH$ lattice from an absolutely distinct perspective of band-topology analysis. Moreover, the localization dependency of light-states on topological parameter of the $AAH$ lattice has analytically been investigated by using the tight-binding lattice approximation, which has also been verified numerically in the proposed waveguide lattice for an appropriate choice of topological parameter. Here, we have revealed that to strongly localize a specific light-state, having higher eigenenergy, higher value of quasi-periodic modulation strength is imperative. We have also reveled that, in the nontrivial topological condition of the optical lattice (i.e., after the topological phase transition), the input excitations in the form of $LE$ ,$ME$ and $RE$ will achieve localized forms in distinctly different ways compared to each other for a fixed topological parameter. Furthermore, it has exclusively been observed that the presence of any off-diagonal disorder favours the topology induced localization phenomenon in the proposed $AAH$ model based optical lattice due to inherent topological protection. Hence, the investigation shows unconventional light dynamics of AAH model based optical lattices and reveals detailed understanding of the effect of topology on light localization. It would open up a new paradigm in the domain of topological photonics by controlling topological features of quasi-periodic structure judiciously to manipulate the exclusive properties of robust state of light in varieties of potential topological devices. 

\section{Acknowledgements}  

SD acknowledges the financial support from the Council of Scientific \& Industrial Research (CSIR), India. The authors thank Mr. Arnab Laha for helpful discussions. 
SG acknowledges financial support from Science and Engineering Research Board (SERB) [Grant No. ECR/2017/000491], India

\bibliography{Ref}

\end{document}